\documentclass[pra,twocolumn,showpacs,groupedaddress,superscriptaddress,aps,10pt]{revtex4-1}
\usepackage{bm,graphicx,amsmath}
\usepackage{placeins}
\usepackage{amssymb}
\usepackage{amsmath}
\usepackage{epsfig}
\usepackage{amssymb}
\usepackage{color}
\usepackage[colorlinks,linkcolor=blue,citecolor=blue]{hyperref}
\usepackage{subfigure}

\begin{document}

\title{Transformation of vector beams with radial and azimuthal polarizations in biaxial crystals}
\date{\today}

\author{Alex Turpin} \email{Corresponding author: alejandro.turpin@uab.cat}
\affiliation{Departament de F\'isica, Universitat Aut\`onoma de Barcelona, Bellaterra, E-08193, Spain}
\author{Asticio Vargas}
\affiliation{Departamento de Ciencias F\'isicas, Universidad de La Frontera, Temuco, Casilla 54-D, Chile}
\affiliation{Center for Optics and Photonics, Universidad de Concepci\'on, Casilla 4012, Concepci\'on, Chile}
\author{Angel Lizana}
\affiliation{Departament de F\'isica, Universitat Aut\`onoma de Barcelona, Bellaterra, E-08193, Spain}
\author{Fabi\'an A. Torres-Ruiz}
\affiliation{Departamento de Ciencias F\'isicas, Universidad de La Frontera, Temuco, Casilla 54-D, Chile}
\affiliation{Center for Optics and Photonics, Universidad de Concepci\'on, Casilla 4012, Concepci\'on, Chile}
\author{Irene Est\'evez}
\affiliation{Departament de F\'isica, Universitat Aut\`onoma de Barcelona, Bellaterra, E-08193, Spain}
\author{Ignacio Moreno}
\affiliation{Departamento de Ciencia de Materiales, \'Optica y Tecnolog\'ia Electr\'onica, Universidad Miguel Hern\'andez, 03202 Elche, Spain}
\author{Juan Campos}
\affiliation{Departament de F\'isica, Universitat Aut\`onoma de Barcelona, Bellaterra, E-08193, Spain}
\author{Jordi Mompart}
\affiliation{Departament de F\'isica, Universitat Aut\`onoma de Barcelona, Bellaterra, E-08193, Spain}

\begin{abstract}
We present both experimentally and theoretically the transformation of radially and azimuthally polarized vector beams when they propagate through a biaxial crystal and are transformed by the conical refraction phenomenon. We show that, at the focal plane, the transverse pattern is formed by a ring-like light structure with an azimuthal node, being this node found at diametrically opposite points of the ring for radial/azimuthal polarizations. We also prove that the state of polarization of the transformed beams is conical refraction-like, i.e. that every two diametrically opposite points of the light ring are linearly orthogonally polarized.

\textbf{OCIS:} (260.5430) Polarization; (260.1440) Birefringence; (260.1180) Crystal optics
\end{abstract}

\date{\today }
%
%
%
\maketitle
\section{Introduction}

Vortex beams with radial and azimuthal polarizations are the paradigm of vector beams \cite{zhan:2009:aop}. The nature of the electric field make them ideal candidates for applications where extremely tightly focused beams are needed, such as in material processing \cite{wieslaw:2011:prl} or super-resolution imaging \cite{biss:2006:ao}. However, the propagation of radially and azimuthally vortex beams through anisotropic media such as uniaxial or biaxial crystals remains almost unexplored \cite{italy:2003:oc,wang:2008:ao,sato:2008:josaa,tan:2010:epj,fedeyeva:2010:josaa}.

\begin{figure}[htb]
\centering
\includegraphics[width=1 \columnwidth]{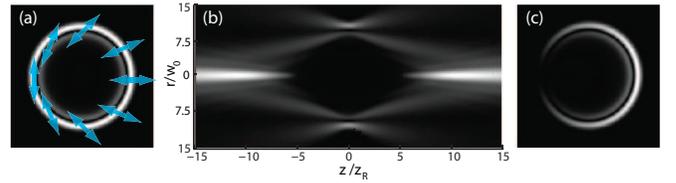}
\caption{(Color online) Intensity distribution of a CR beam under conditions of $\rho_0 \equiv R_0 / w_0 \gg 1$. (a) Bright and dark Poggendorff rings at the focal plane obtained with a circularly polarized Gaussian input beam. Blue double-arrows indicate the polarization distribution along the rings. (b) Intensity distribution in the $z$-$r$ plane showing the free space evolution of the CR beam. (c) CR transverse intensity pattern at the focal plane for a horizontally polarized Gaussian input beam.}
\label{fig1}
\end{figure}

In anisotropic crystals, the most particularly spectacular phenomenon is conical refraction (CR), occurring  in biaxial crystals \cite{belsky:1978:os,berry:2004:jo,kalkandjiev:2008:spie}. In CR, a homogeneously circularly polarized focused Gaussian beam passing along one of the optic axis of a biaxial crystal is transformed into a pair of concentric bright rings with a dark (Poggendorff) ring, at the focal plane of the system \cite{belsky:1978:os,berry:2004:jo,kalkandjiev:2008:spie,peet:2010:oc}, see Fig.~\ref{fig1}(a). The state of polarization (SOP) of the rings is the following: any point of the rings is linearly polarized with the azimuth $\Phi$ rotating along the ring so that every two diametrically points are orthogonally polarized, i.e. $\Phi = \frac{\varphi - \varphi_C}{2}$, where $\varphi$ is the azimuth in cylindrical coordinates and $\varphi_C$ is the orientation of the plane of optic axes of the crystal \cite{kalkandjiev:2008:spie,turpin_ebs:2013:oe}. By moving symmetrically from the focal plane in the axial direction, the CR transverse pattern becomes more involved and two intensity peaks, known as Raman spots \cite{raman:1941:nature}, are found at axial positions $z$ given by $z_{\rm{Raman}} = \pm \sqrt{\frac{4}{3}} \rho_0 z_{R}$ \cite{turpin_vault:2013:oe}, where $z_R$ is the Rayleigh range of the focused input beam. Therefore, the CR beam forms an optical bottle closed axially by the two Raman spots, as shown in Fig.~\ref{fig1}(b). Note that all these features of the CR beam are found for a circularly polarized input beam of Gaussian transverse profile satisfying the condition $\rho_0 \equiv R_0/w_0 \gg 1$, which covers most of the applications of CR \cite{phelan:2010:oe,turpin:2012:ol,peinado:2013:ol,turpin_vault:2013:oe,turpin_bec:2015:oe}. For a linearly polarized input beam, the CR ring possesses an azimuthal node corresponding with the polarization of the ring with opposite azimuth to the input beam, see Fig.~\ref{fig1}(c). 

CR related to radially and azimuthally polarized beams, in what follows named as R/A beams, has been studied recently showing that R/A beams can be obtained from CR by using an interferometric system \cite{phelan:2011:oe}, by using linearly polarized Bessel beams \cite{khilo:2012:oc} and from the transformation of input R/A vortex beams by means of CR \cite{amin:2014:ol}. In this work, we use a similar experimental set-up as in \cite{amin:2014:ol} to show the beam evolution and SOP of the CR beams obtained beyond the biaxial crystal when input R/A vector beams are used (Section~\ref{sec2}). In Section~\ref{sec3} by means of the diffractive theory of CR \cite{belsky:1978:os,berry:2004:jo,turpin_general:2015:arxiv} we perform the numerical simulations corresponding to our experimental scheme, finding a reasonable agreement between both numerical simulations and experimental results. Finally, arguments to understand the origin of the differences between our results and the ones reported in \cite{amin:2014:ol} and the summary of our work are provided in Section~\ref{sec4}.

\section{Experimental procedure}
\label{sec2}

\subsection{Experimental set-up}

\begin{figure}[htb]
\centering
\includegraphics[width=0.9 \columnwidth]{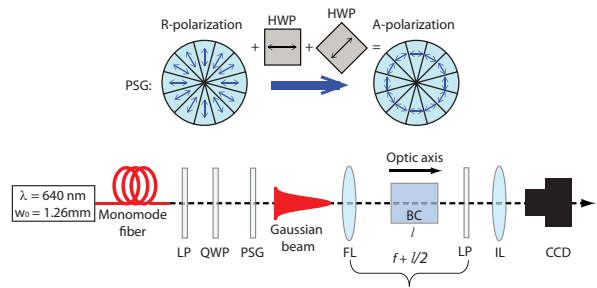}
\caption{(Color online) Experimental set-up. A circularly polarized Gaussian beam is obtained from a diode laser couple to a mono-mode fiber, a linear polarizer (LP) and a quarter waveplate (QWP). A radially (R) polarized beam is generated when the Gaussian beam passes through a segmented radial polarizer. To obtain an azimuthally (A) polarized beam, we use additionally two half waveplates with $45^{\circ}$ rotation of the mutual orientation of their axis. The R/A beam is then focused by a lens (FL, $200\,\rm{mm}$ of focal length) being its propagation direction paralel to one of the optic axes of a biaxial crystal (BC) of length $l=23\,\rm{mm}$. An additional LP is used to check the state of polarization of the output beam. Finally, an imaging lens (IL) is used to project the transverse pattern along the axial direction into the CCD camera.}
\label{fig2}
\end{figure}

The experimental set-up is sketched in Fig.~\ref{fig2}. A diode laser at $640\,\rm{nm}$ coupled to a monomode fiber generates a Gaussian beam that passes through a linear polarizer (LP), a quarter waveplate (QWP) to ensure circular polarization and a polarization state generator (PSG). The PSG consists in 12 linear polarizers joint to form an azimuthally segmented polarizer that transforms the input polarization into radial, see top part of Fig.~\ref{fig2}. Therefore, after passing through the PSG a Gaussian R-polarized beam is obtained. To generate an A-polarized beam, two half wave-plates (HWPs) with a mutual angle of $45^{\circ}$ are additionally arranged into the system. Unfortunately, the two HWPs used do not operate perfectly for the wavelength used and we have checked that the A-polarized beam obtained has an ellipticity of nearly $3^{\circ}$. The R/A beam is focused with a lens of $200\,\rm{mm}$ focal length and passes through a biaxial crystal and parallel to one of its optic axis. The biaxial crystal is placed always before the expected focal plane of the beam in the absence of the crystal. An additional imaging lens is used to transfer the transverse planes along the beam propagation direction onto the CCD camera. A LP is used as analyzer, to check the state of polarization of the beam after being transformed by the CR phenomenon. We use a commercially available (CROptics) $\rm{KGd(WO_4)_2}$ biaxial crystal with $\alpha = 16.9$~mrad and length $l = 23\,\rm{mm}$ yielding CR ring radius of $R_0 \approx 390\,\mu\rm{m}$. For more details about the material and the alignment procedure see e.g. Ref.~\cite{kalkandjiev:2008:spie}. 

\subsection{Experimental results}

\begin{figure}[htb]
\centering
\includegraphics[width=0.9 \columnwidth]{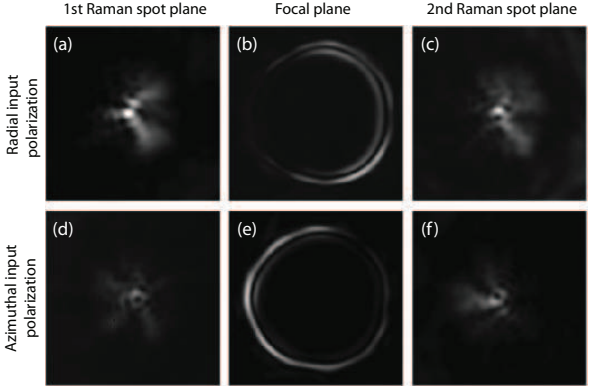}
\caption{Experimental transverse intensity patterns obtained from a R/A-polarized input beam (first/second row) at (a,d) the first Raman spot plane, (b,e) the focal plane and (c,f) the second Raman spot plane.}
\label{fig3}
\end{figure}

Fig.~\ref{fig3} presents the experimental transverse intensity patterns at the focal plane an at the Raman spots obtained from a R-polarized (first row) and an A-polarized (second row) input beam, in the absence of the linear analyzer in Fig.~\ref{fig2}. At the focal plane and for both input polarizations the transverse pattern is analogous to what would be obtained in case of a linearly polarized input beam: the intensity of the CR rings is azimuthally crescent, with a point of null intensity (in the case of an input A-polarized beam, i.e. Fig.~\ref{fig3}(e), this null intensity becomes a region of minimum intensity due to the non-perfect operation of the half waveplates) . However, by comparing Figs.~\ref{fig3}(b) and (e) it can be observed that the node for an A-polarized input beam appears at a diametrically opposite position along the ring with respect to a R-polarized input beam. 
Note that the maximum intensity points on the transverse patterns at the focal plane are found at the position of the azimuthal sector of the R/A-polarized input beam coinciding with the CR polarization. In contrast, the minimum intensity points are found at positions where the R/A-polarized input beam is orthogonal to the CR polarization.

At the position of the Raman spots and for both R/A-polarized beams, the transverse pattern is similar to the one expected for a uniformly linearly polarized Gaussian input beam, with the difference that now a node is formed at the beam center. This is expected, since at that point there is a polarization singularity due to nature of the R/A-polarized beams. 
The reported experimental transverse intensity patterns are not perfectly uniform due to the discreteness of the PSG used to obtain the R/A-polarized input beams, which is made of 12 individual polarizers joint together.

\begin{figure}[htb]
\centering
\includegraphics[width=1 \columnwidth]{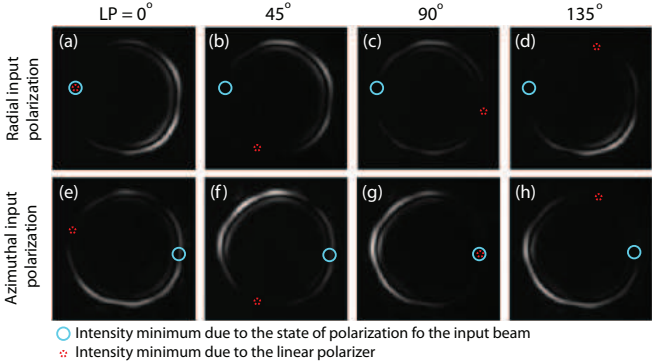}
\caption{(Color online) Experimental transverse intensity patterns obtained from a R/A-polarized input beam (first/second row) at the focal plane when a linear polarizer (LP) used as analyzer after the crystal is oriented at (a,e) $0^{\circ}$, (b,f) $45^{\circ}$, (c,g) $90^{\circ}$ and (d,h) $135^{\circ}$.}
\label{figpol}
\end{figure}

Fig.~\ref{figpol} shows the transverse patterns of the CR beams at the focal plane when a linear polarizer (LP) is used as analyzer and rotated at angles $[0,45,90,135]^{\circ}$. Both for R- and A-polarized beams, the node introduced by the LP appear at the same azimuthal position. Additionally, the LP only introduces one node, being this placed at opposite positions along the ring when comparing the results corresponding to $0^{\circ}$ and $90^{\circ}$ [Figs.~\ref{figpol}(a,e) and (c,g), respectively], and $45^{\circ}$ and $135^{\circ}$ [Figs.~\ref{figpol}(b,f) and (d,h), respectively]. Therefore, we can conclude that the state of polarization of the CR is the same for both R/A input polarizations and that it is CR-like. 

\section{Theoretical model}
\label{sec3}
In order to corroborate to experimental results presented above, we have performed the corresponding numerical simulations of transformation of R/A-polarized Gaussian beams by means of CR. We consider R/A-polarized input beams with transverse electric field 
\begin{eqnarray}
\vec{E}_{R} = e^{-r^2} (\cos\phi,\sin\phi) e^{i\phi},\label{E_rad} \\ 
\vec{E}_{A} = e^{-r^2} (\sin\phi,-\cos\phi) e^{i\phi}, \label{E_azi}
\end{eqnarray}
where $r = \sqrt{x^2+y^2}$ and $\phi = \tan^{-1}(y/x)$; and Fourier transform $\vec{A}(\vec{\kappa})$
\begin{eqnarray}
\vec{A}(\vec{\kappa}) &=& A_x(\vec{\kappa}) \vec{e_x} + A_y(\vec{\kappa}) \vec{e_y}, \\ \label{FT_vector}
A_{x}(\vec{\kappa}) &=& \frac{1}{(2 \pi)^2} \iint\limits_{-\infty}^{\infty}
E_{x}(\vec{r})  e^{-i \vec{\kappa} \cdot \vec{r}} dx dy,\\ \label{FT_x}
A_{y}(\vec{\kappa}) &=& \frac{1}{(2 \pi)^2} \iint\limits_{-\infty}^{\infty}
E_{y}(\vec{r})  e^{-i \vec{\kappa} \cdot \vec{r}} dx dy, \label{FT_y}
\end{eqnarray}
where $\vec{\kappa} = (\kappa_x,\kappa_y)$. Note that the spiral phase term ($\exp(i \phi)$) in Eqs.~(\ref{E_rad}) and (\ref{E_azi}) comes from the generation of the R/A-polarization with the segmented polarizer illuminated with circular polarization.
Beyond the crystal, the electric field of the transformed beam is given by the following set of equations \cite{turpin_general:2015:arxiv}:
%
\begin{eqnarray}
B_{0, \alpha}(\vec{r},\rho_0)= 
\iint\limits_{-\infty}^{\infty} \frac{i e^{-i \Phi}}{(2 \pi)^2} \frac{\kappa_y}{\kappa} \sin \left( \rho_0 \kappa \right) A_{\alpha}(\vec{\kappa}) d \vec{\kappa},\label{B0_CR} \\ 
B_{1, \alpha}(\vec{r},\rho_0)=
\iint\limits_{-\infty}^{\infty} \frac{e^{-i \Phi}}{(2 \pi)^2} \left( \cos \left( \rho_0 \kappa \right) + i \frac{\kappa_x}{\kappa} \sin \left( \rho_0 \kappa \right) \right) A_{\alpha}(\vec{\kappa}) d \vec{\kappa},~\label{B1_CR}
\end{eqnarray}
where $\alpha=\{x,y\}$, $n$ is the mean refractive index of the biaxial crystal, $\vec{r} = (x,y)$, $\kappa = \sqrt{\kappa_x^2 + \kappa_y^2}$ and $\Phi = \vec{\kappa} \cdot \vec{r} - \frac{z}{2n} \kappa^2$. The expressions for the displacement vector $\vec{D}$ in terms of Eqs.~(\ref{B0_CR}) and (\ref{B1_CR}) are
\begin{eqnarray}
D_{x} &=& B_{0, y}(\vec{r},\rho_0) + B_{1, x}(\vec{r},\rho_0),
\label{E_CRx}\\
D_{y} &=& B_{0, x}(\vec{r},\rho_0) + B_{1, y}(\vec{r},-\rho_0). 
\label{E_CRy}
\end{eqnarray}

\begin{figure}[htb]
\centering
\includegraphics[width=1 \columnwidth]{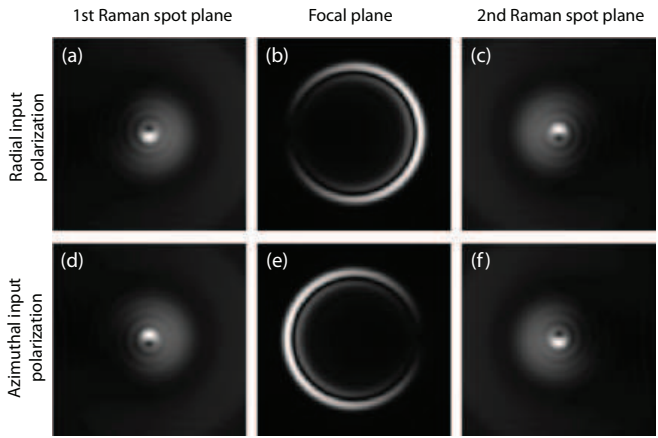}
\caption{Numerically calculated transverse intensity patterns obtained by using Eqs.~(\ref{E_rad})--(\ref{E_CRy}) from a R/A-polarized input beam (first/second row) at (a,d) the first Raman spot plane, (b,e) the focal plane and (c,f) the second Raman spot plane.}
\label{fig5}
\end{figure}

\begin{figure}[htb]
\centering
\includegraphics[width=1 \columnwidth]{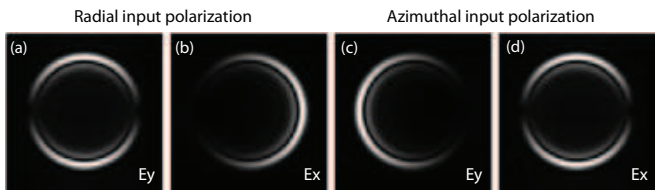}
\caption{Numerically calculated transverse intensity patterns of the (a,c) $E_y$ and (b,d) $E_x$ fields obtained by using Eqs.~(\ref{E_rad})--(\ref{E_CRy}) from a R/A-polarized input beam (a,b/c,d) at the focal plane.}
\label{fig6}
\end{figure}

Fig.~\ref{fig5} presents the numerically calculated transverse intensity pattern of the CR beam for radial and azimuthal input polarizations. The corresponding electric field in the $x$ and $y$ direction of at the focal plane is shown in Fig.~\ref{fig6}.
As it can be appreciated, the numerical simulations obtained with Eqs.~(\ref{E_rad})--(\ref{E_CRy}) reproduce the experimental observations shown above: (i) the transverse pattern at the focal plane is formed by an annular light structure with an azimuthal node, (ii) the position of the node is at diametrically opposite points when R/A-polarized beams are used, (iii) the state of polarization at the focal plane is CR-like, i.e. every two diametrically opposite points are orthogonally polarized; (iv) far from the focal plane a vortex-like beam is observed at the beam center and (vi) the beam evolution is non symmetric with respect to the focal plane, at variance with CR beams obtained from homogeneously polarized Gaussian input beams. 

\section{Discussion and conclusions}
\label{sec4}
In this section we discuss the similarities and differences of our results with the work of Grant \textit{et al} \cite{amin:2014:ol}. On the one hand, the transverse pattern that we have found at the focal plane for R/A-polarizations is the same as in Ref.~\cite{amin:2014:ol}. On the other hand, we have found additionally a vortex-like structure far from the focal plane, at variance with what was reported in Ref.~\cite{amin:2014:ol}. Note that a R/A-polarized beam possesses an optical singularity in the electric field at the beam center and this optical singularity must be preserved along its transmission through a linear system, such as a biaxial crystal, as we have found. Additionally and what is more important, the state of polarization of the CR pattern at the focal plane that we have found is CR-like. In contrast, in Ref.~\cite{amin:2014:ol} CR patterns with R/A-polarizations were reported. As shown in \cite{phelan:2009:oe,turpin_ebs:2013:oe,peinado:2013:ol,turpin_stokes:2015:oe}, the polarization distribution at the focal plane for $\rho_0 \gg 1$ is always linear at any point of the ring being every pair of diametrically opposite points orthogonally polarized with independence of the state of polarization of the input beam. Therefore, even for non-homogeneously R/A-polarized beam, the polarization distribution of the CR rings at the focal plane must be CR-like. 

Finally, note that in our case the input beam is R/A-polarized with a spiral phase term due to the helicity of the circularly polarized Gaussian beam passing through the segmented polarizer. In contrast, in Ref.~\cite{amin:2014:ol} the input beam passing through the biaxial crystal is a perfect R/A-polarized beam, since it is generated with a S-plate. However, we have numerically checked that all the differences commented above with respect to the features of the CR beams reported in Ref.~\cite{amin:2014:ol} are still found, see Appendix~\ref{App:AppendixA}.

To sum up, we have reported the transformation of radially and azimuthally polarized beams by a biaxial crystal throughout the conical refraction phenomenon and we have obtained a reasonable agreement between experiments and numerical simulations. We have demonstrated that the transverse intensity pattern at the focal plane is formed by two concentric light rings with an azimuthal node. The state of polarization of these rings has been shown to be CR-like, i.e. such that every two opposite points of the rings are orthogonally polarized. Finally, we have found that far enough from the focal plane along the axial direction a vortex-like beam is obtained and a non-symmetric beam evolution. 

\section*{Acknowledgments}

Spanish Ministry of Science and Innovation (MICINN) (contracts FIS2011-23719 and FIS2012-39158-C02-01, fondos FEDER and grant AP2010-2310). 
Catalan government (contract 2014SGR1639).
Chilean Ministry of Science (contracts PIA-CONICYT PF0824, FONDECYT 1151290 and FONDECYT 11110258). 
Dr. T. K. Kalkandjiev is acknowledged for supplying the biaxial crystal used in the experiments.

\appendix
\section{Numerical calculations for radially and azimuthally polarized beams obtained from a S-plate} 
\label{App:AppendixA}
\begin{figure}[htb]
\centering
\includegraphics[width=1 \columnwidth]{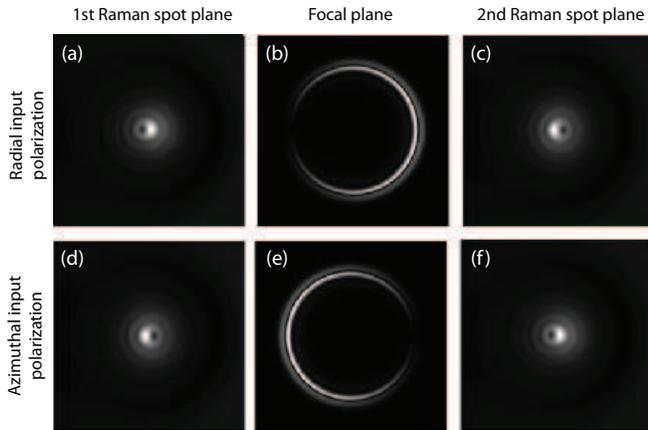}
\caption{Numerically calculated transverse intensity patterns obtained by using Eqs.~(\ref{FT_vector})--(\ref{E_azi_s}), from a R/A-polarized input beam (first/second row) obtained with a S-plate at (a,d) the first Raman spot plane, (b,e) the focal plane and (c,f) the second Raman spot plane.}
\label{fig7}
\end{figure}

\begin{figure}[htb]
\centering
\includegraphics[width=1 \columnwidth]{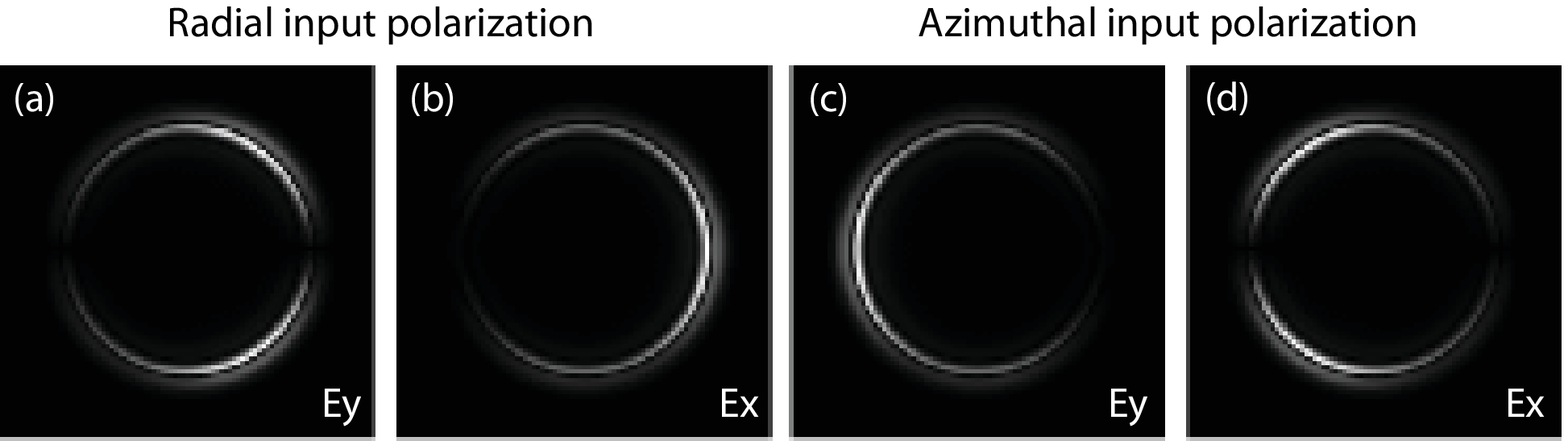}
\caption{Numerically calculated transverse intensity patterns of the (a,c) $E_y$ and (b,d) $E_x$ fields obtained by using Eqs.~(\ref{FT_vector})--(\ref{E_azi_s}) from a R/A-polarized input beam (a,b/c,d) obtained with a S-plate at the focal plane.}
\label{fig8}
\end{figure}

In this Appendix we present the transverse intensity patterns for the electric field behind the biaxial crystal obtained when radially and azimuthally (R/A) polarized beams generated with a S-plate are considered. Such beams can be described as follows:
\begin{eqnarray}
\vec{E}_{R}' = e^{-r^2} (\cos\phi,\sin\phi),\label{E_rad_s} \\ 
\vec{E}_{A}' = e^{-r^2} (\sin\phi,-\cos\phi). \label{E_azi_s}
\end{eqnarray}
Fig.~\ref{fig7} presents the numerically calculated transverse intensity pattern of the CR beam for R/A input polarizations. The corresponding electric field in the $x$ and $y$ direction of at the focal plane is shown in Fig.~\ref{fig8}. As it can be appreciated, the polarization distribution of the CR beam at the focal plane is as reported in Fig.~\ref{fig6}, i. e. CR-like. Additionally, a node at the Raman spot can be also found, as in the case of R/A-polarized beams with a spiral phase term, see Fig.~\ref{fig5}. 

\end{document}